\documentclass[slac_one]{revtex4}
\usepackage{graphicx}
\usepackage{fancyhdr}
\def\ie{{\it i.e.}}
\def\eg{{\it e.g.}}

\def\mpl{\ifmmode M_{pl}\else $M_{pl}$\fi}
\def\mpl{\ifmmode \overline M_{Pl}\else $\bar M_{Pl}$\fi}

\def\atversim#1#2{\lower0.7ex\vbox{\baselineskip\zatskip\lineskip\zatskip
  \lineskiplimit 0pt\ialign{$\matth#1\hfil##\hfil$\crcr#2\crcr\sim\crcr}}}

\begin{document}
\rightline{\vbox{\halign{&#\hfil\cr
&SLAC-PUB-13370\cr
}}}

\title{$Z'$ Physics at the LHC and the LHeC}

\author{Thomas ~G.~Rizzo}
\affiliation{SLAC, Stanford, CA 94025, USA}

\begin{abstract}
The methods which can be employed to determine the properties of new neutral resonant states that may be observed 
in the Drell-Yan channel at the LHC are reviewed. If these states are sufficiently light we discuss how polarized 
$ep$ scattering at the LHeC can assist in the determination of their couplings to the Standard Model (SM) fermions. 
\end{abstract}

\maketitle

\thispagestyle{fancy}

\begin{figure}[htbp]
\includegraphics[width=5.0cm,angle=90]{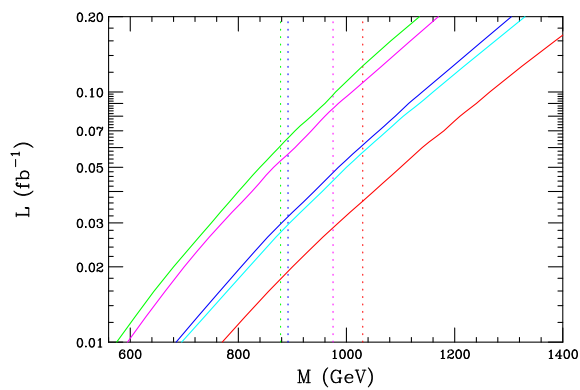}
\vspace*{-0.1cm}
\caption{$5\sigma$ discovery reach for familiar $Z'$ models at the 10 TeV LHC as a function of the integrated 
luminosity (solid curves) and the corresponding present limits from CDF{\cite {TeV_ICHEP}} (dotted lines). The 
red(green,blue,magenta,cyan) lines correspond to the SSM($\psi$,$\chi$,$\eta$,LRM) cases}
\label{fig1}
\end{figure}

A TeV-scale $Z'$-like object is a common prediction of many beyond the SM scenarios{\cite {classic,new}}. Extended 
gauge models, R-parity violating SUSY, string constructions/intersecting brane models, Little Higgs scenarios, 
Hidden Valley models, Kaluza-Klein excitations of the gauge and gravity sectors in models with extra dimensions, 
string excitations and unparticles all provide such examples. One of the most exciting aspects of these states 
is that they can lead to large signals in the experimentally clean Drell-Yan channel at the LHC and thus may be the 
first signal for new physics to be observed. While it is well known that the eventual mass reach for such states at 
the LHC is in the multi-TeV range, even the low luminosity run at $\sqrt s=10$ TeV has a window for discovery while 
still satisfying the constraints from the Tevatron{\cite {TeV_ICHEP}}. This can be seen explicitly in Fig.~\ref{fig1} 
for the familiar $E_6$-type $Z'$ models{\cite {classic}}. Generally such a discovery only requires a few handfuls of 
events. 

If a $Z'$-like is state is discovered at the LHC we will want to know all about it. The most important things to learn 
first are its `line-shape' parameters, \ie, the mass ($M$), width ($\Gamma$) and leptonic production cross section 
($\sigma$). It is important to note that for some models, such as in the case of unparticles, this line shape may not 
even be of the usual Breit-Wigner form{\cite {me2}}. The next thing to learn is the spin of the resonance; R-parity 
violating sneutrinos and Kaluza-Klein gravitons are spin-0 and spin-2 resonances, respectively, whereas most of the 
other possibilities are spin-1. To determine the spin of this resonance requires a measurement of the decay angular 
distribution mandating at least a few hundred events; clear spin determinations will be difficult for very massive 
states. Note that if the $\gamma \gamma$ final state is observed in the decay of the resonance we will know immediately 
that it is not spin-1. 

Once we know that we have discovered a spin-1 $Z'$ we will want to determine its couplings to the SM fermions. How many 
independent such couplings are there? In the simplest case, these couplings will be generation independent. Furthermore, 
in many models, the generator associated with the $Z'$ will also commute with the SM weak isospin. If these two 
conditions are satisfied then there are only 5 independent couplings to determine, one for each of the SM representations: 
$Q,L,u^c,d^c,e^c$. Can these 5 parameters be uniquely determined by the LHC at full integrated luminosity, \ie, 100 
$fb^{-1}$?. Unfortunately, the answer to this question is `No', but several combinations{\cite {TAIT}} of these couplings can 
be determined with a respectable precision even when NLO QCD corrections are included{\cite {FRANK}}. Although this may help us 
to differentiate between various models it does not provide us with all of the information we need for complete coupling 
determination.

\begin{figure}[htbp]
\includegraphics[width=8.0cm,angle=0]{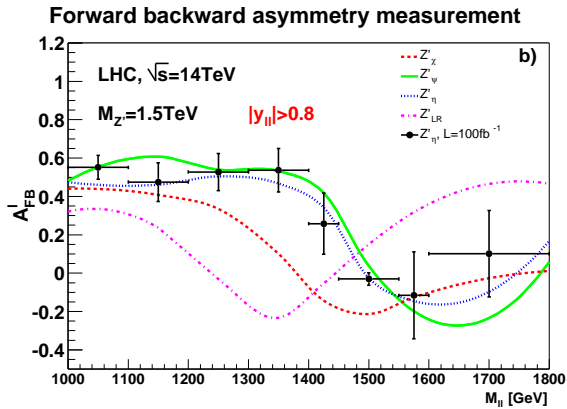}
\vspace*{-0.1cm}
\caption{Sample $A_{FB}$ measurements in the dilepton mass region near a $Z'$.}
\label{fig2}
\end{figure}

What observables are available to make such determinations and can we do it in as model-independent a fashion as possible, \eg, 
can we perform this analysis without the assumption that the $Z'$ decays only to SM particles{\cite {classic,new}}? Clearly, \eg, 
the values of both $\sigma$ and $\Gamma$ depend upon other decay modes being present. However, one can show that the product, $\sigma \Gamma$, is 
only very weakly sensitive to this possibility thus providing us our first useful observable. Like the spin determination, only a few hundred 
evens will be necessary to get a reasonable determination of this quantity. A second well-known observable is the 
forward-backward asymmetry, $A_{FB}$; to measure this observable one must determine the scattering angle between the original 
quark direction and that of the negatively charged lepton in the final state. Unfortunately, the original quark direction can only be 
inferred from the boost of the dilepton system since one expects the average $x$ for valence quarks to be larger than those for sea 
quarks. Since this educated `guess' of the original quark direction will be wrong some of the time, dilution of the asymmetry 
will occur and it must be corrected for using Monte Carlo. Both ATLAS and CMS have shown that these corrections can be done 
successfully to recover the true $A_{FB}$ if a sufficiently strong cut is placed on the dilepton system, \eg, $Y_{ll}\geq 0.8$. 
We can, in principle, obtain two pieces of information from $A_{FB}$ depending upon the dilepton invariant mass interval under 
study: the resonance region around the mass peak and separately in the $Z-Z'$ interference region significantly below the mass peak. 
Given the dilepton rapidity cut and the necessarily detailed measurement of the angular distribution this will require $\sim 10^3$ 
events for an $A_{FB}$ determination in both kinematic regions. These are more easily obtainable in the resonance region where the 
cross section is large; a possible example of these measurements for a light $Z'$ can be seen in Fig.~\ref{fig2} from Ref.{\cite {them}}. 
Clearly these 
measurements for a heavy $Z'$ will be difficult due to the falling statistics and a reasonable determination of $A_{FB}$ may not 
be readily available in these cases.  

Another observable makes use of the dilepton pair distribution directly. At fixed dilepton mass the $u\bar u,d\bar d$ and sea contributions 
lead to different rapidity distributions for the dileptons and the knowledge of the PDFs gives their overall `luminosity' weighting. Thus a measurement 
of the rapidity of the dilepton distribution can provide information on the $u$ and $d$ quark couplings of the $Z'$. As shown in 
Refs{\cite {TAIT,FRANK}}, with $\sim 10^3$ events and by combining with the other observables above one can get a reasonable handle on certain  
combinations of quark couplings. To do better, and requiring $\sim 10^4$ events, one needs SLHC to obtain additional information from $\tau$ 
polarization, ($\gamma,W^\pm,Z)+ Z'$ associated production or a study of rare $Z'$ decay channels{\cite {classic,new}}. It is clear from this discussion 
that this partial information on the $Z'$ couplings can only be obtained for $Z'$ masses $\leq 1.5-2$ TeV unless we go to SLHC luminosities or have 
a TeV-scale linear collider. To get detailed coupling information we will require additional input. 

There has been a proposal either to add an addition $e^\pm$ ring to the LHC or a new linac that will enable us to explore polarized $e^\pm p$ 
interactions in the $\sqrt s=1.5-2$ TeV range{\cite {lhec}}. A discussion of the technicalities of this possibility are beyond the scope of this 
talk but such interactions will allow us an additional probe of the $Z'$ couplings for masses in the above range{\cite {me}} through the 
interference of the $Z'$ with the SM contributions to these amplitudes. Given the four polarized cross sections we can conveniently form six 
polarization-dependent asymmetries in order to reduce PDF and other uncertainties which have varying $Z'$ coupling sensitivities:
\begin{eqnarray}
A^\pm &=& {{d\sigma(e_L^\pm)-d\sigma(e_R^\pm)}\over {d\sigma(e_L^\pm)+d\sigma(e_R^\pm)}}\\ \nonumber
C_{L,R} &=& {{d\sigma(e_{L,R}^-)-d\sigma(e_{L,R}^+)}\over {d\sigma(e_{L,R}^-)+d\sigma(e_{L,R}^+)}}\\ \nonumber
B_{1,2} &=& {{d\sigma(e_{L,R}^-)-d\sigma(e_{R,L}^+)}\over {d\sigma(e_{L,R}^-)+d\sigma(e_{R,L}^+)}}\,.
\end{eqnarray}

\begin{figure}[htbp]
\centerline{
\includegraphics[width=4.5cm,angle=90]{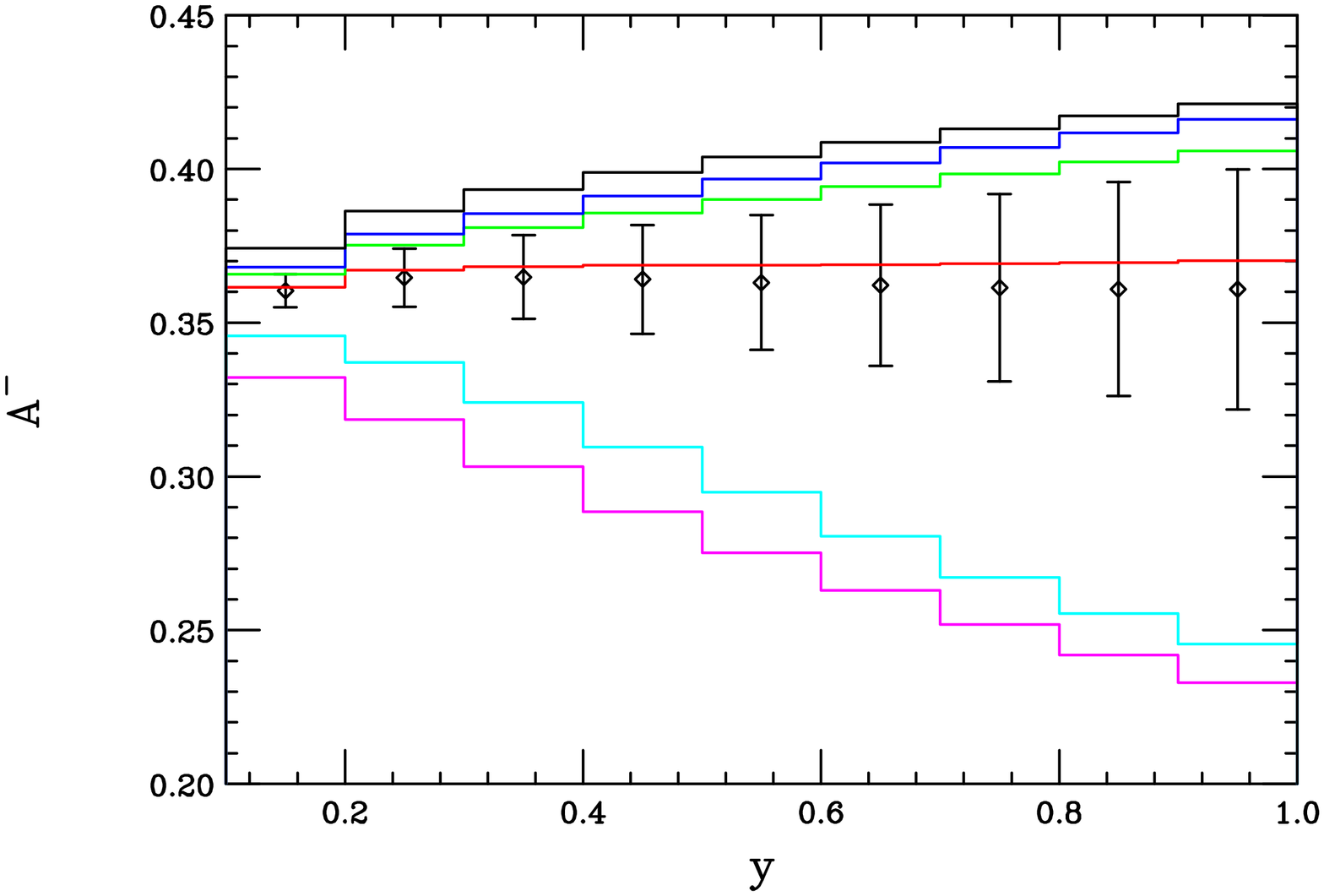}
\hspace*{0.1cm}
\includegraphics[width=4.5cm,angle=90]{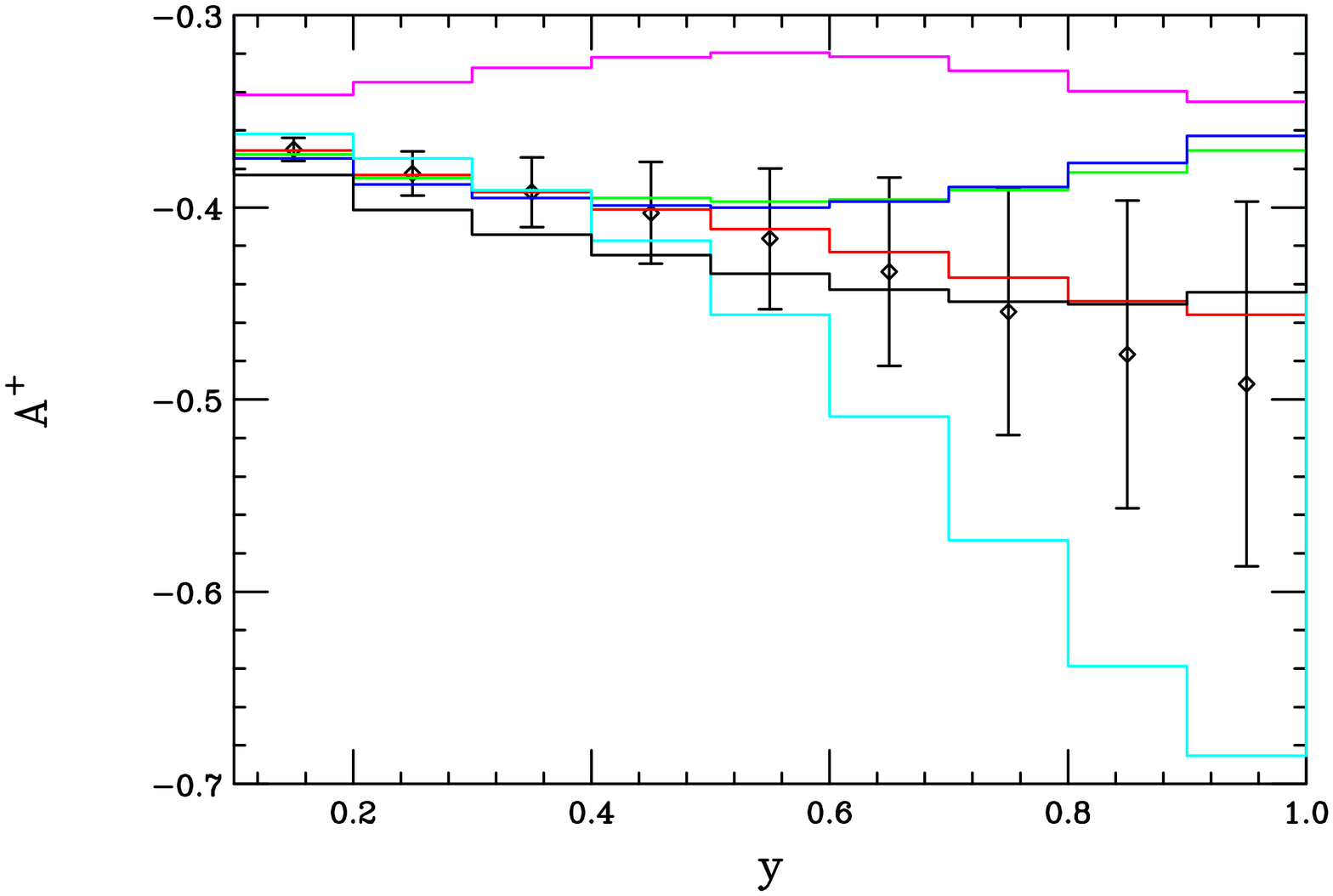}}
\vspace*{0.1cm}
\centerline{
\includegraphics[width=4.5cm,angle=90]{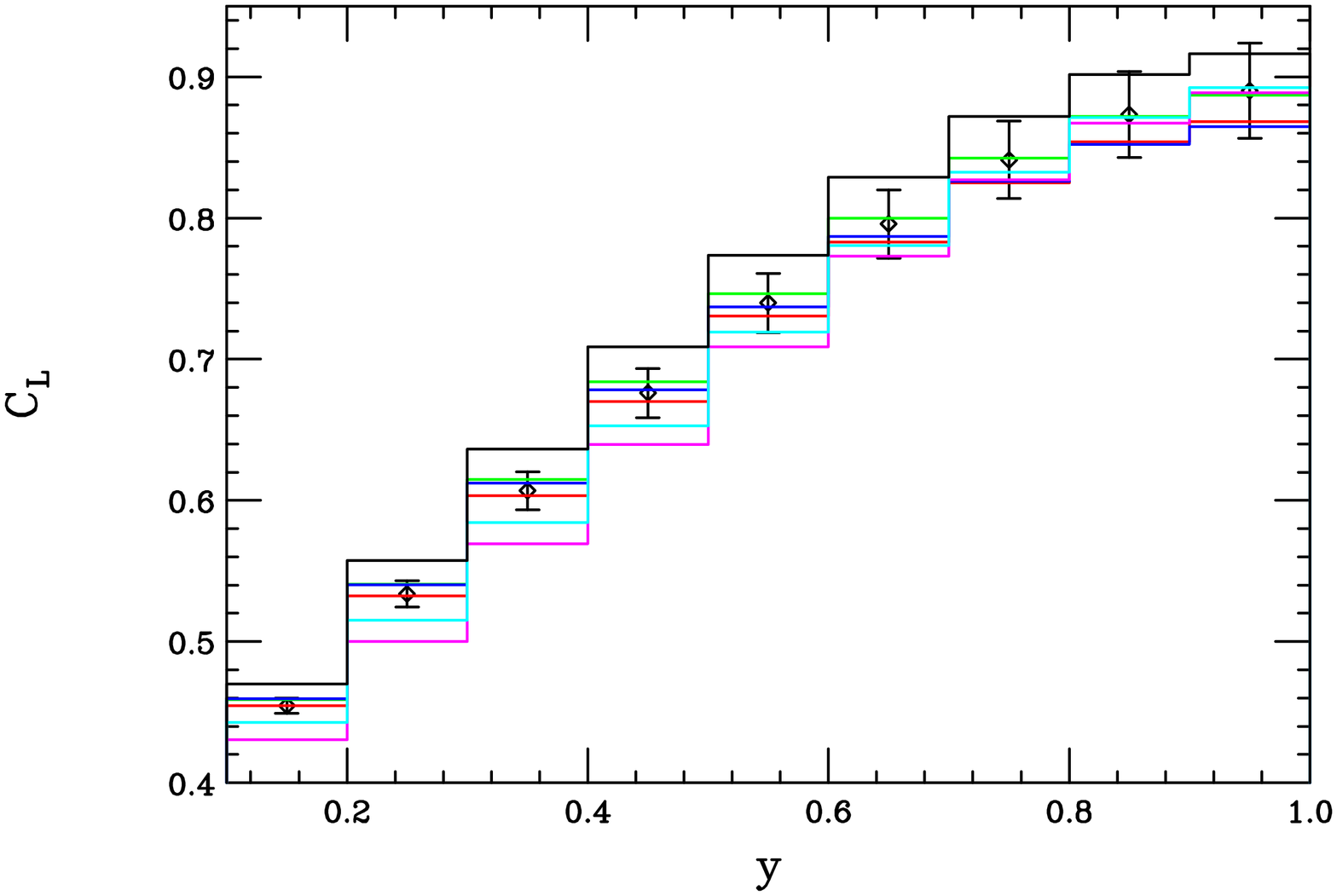}}
\caption{The asymmetries $A^\pm,C_L$ as functions of the DIS scattering variable $y$ in 0.1 bins
as described in the text assuming $\sqrt s=1.5$ TeV and $M_{Z'}=1.2$
TeV. The data points are the SM predictions with their associated errors. The histograms, from top to bottom on the right-hand side of the
top-left (top-right), [bottom] panel correspond to $Z'$ models SSM, $\eta$, $\chi$, $\psi$, ALRM, LRM
(LRM, $\eta$, $\chi$, SSM, $\psi$, ALRM),
[SSM, $\chi$, $\eta$, $\psi$, ALRM, LRM], respectively.}
\label{fig3}
\end{figure}

\begin{figure}[htbp]
\centerline{
\includegraphics[width=4.5cm,angle=90]{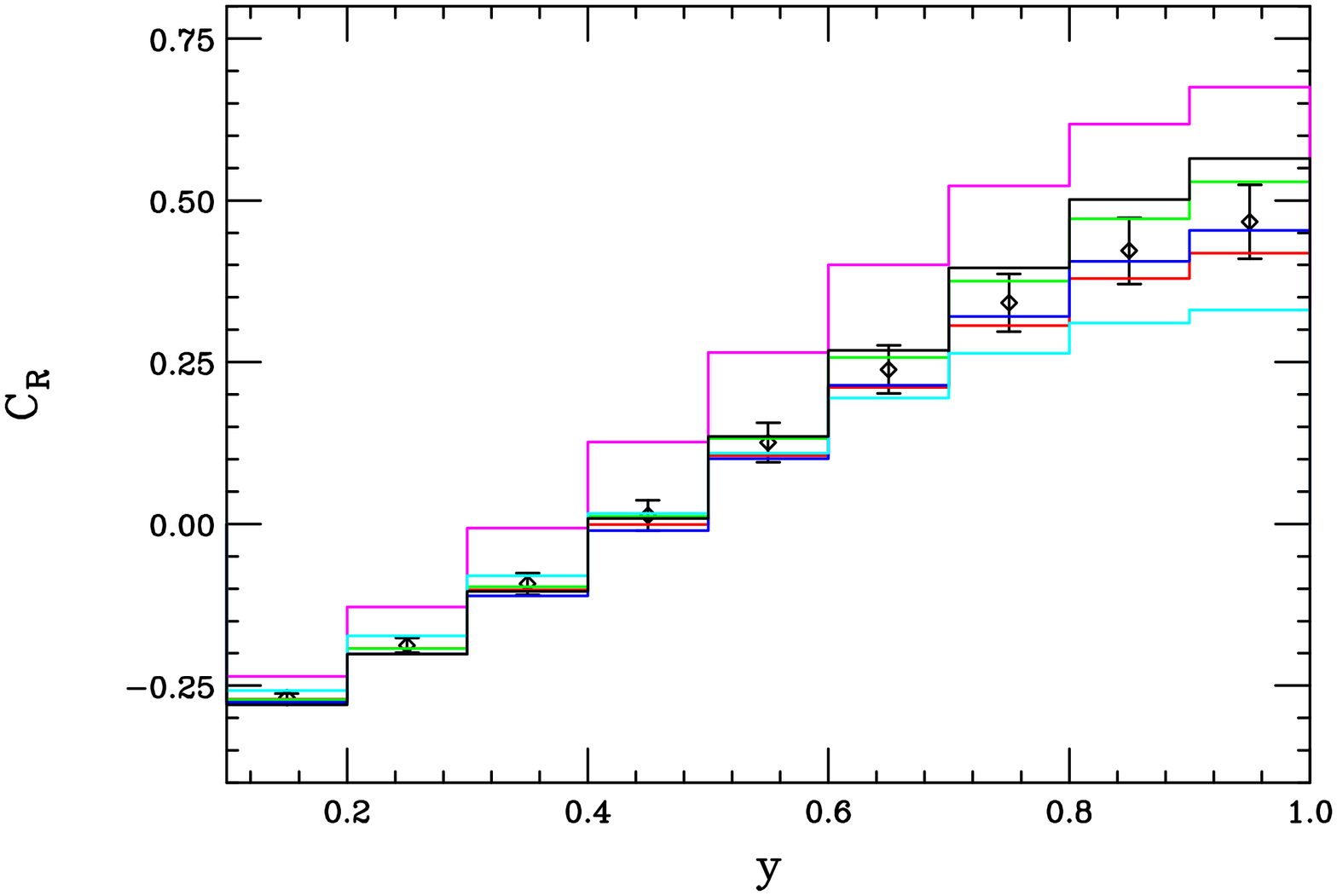}
\hspace*{0.1cm}
\includegraphics[width=4.5cm,angle=90]{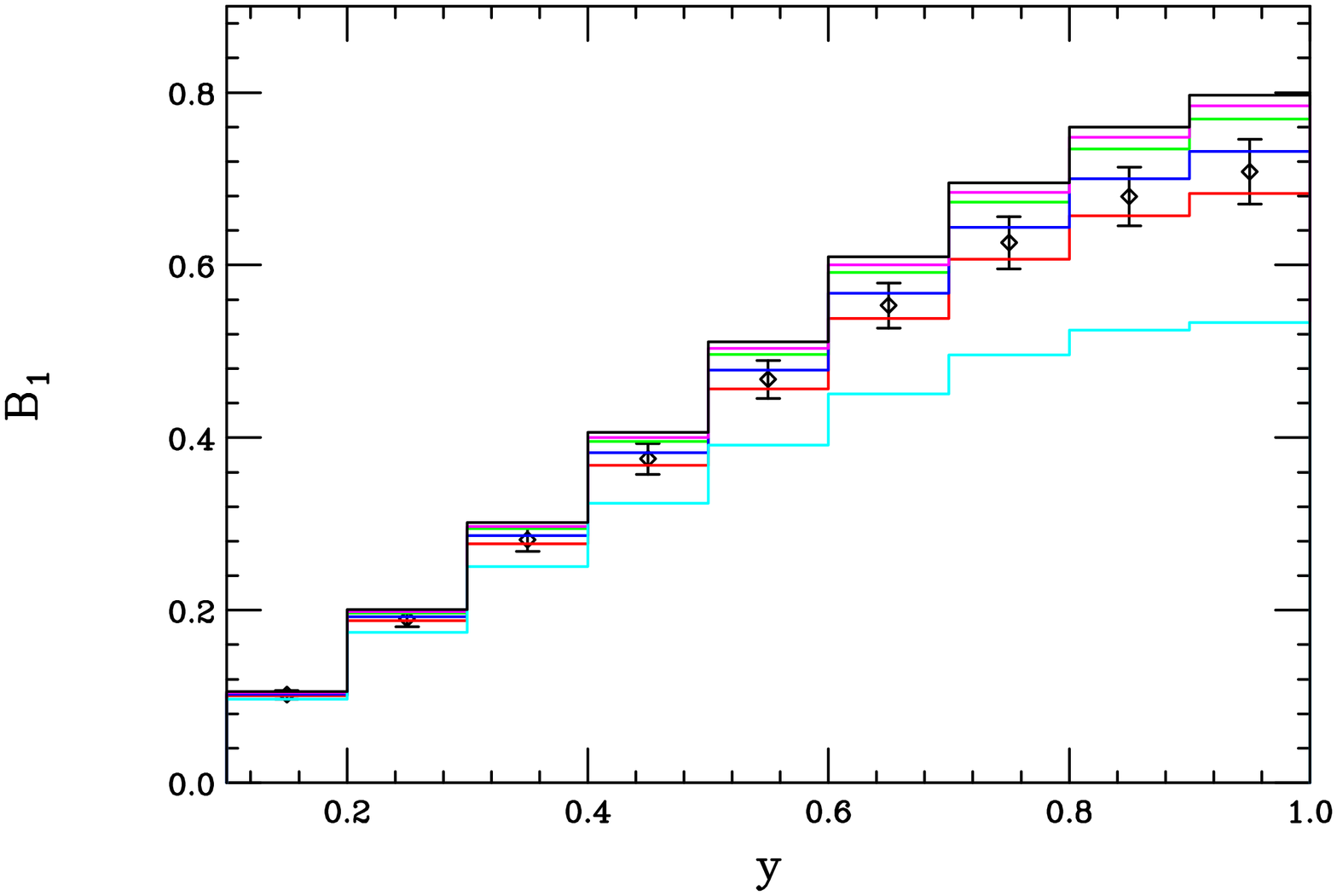}}
\vspace*{0.2cm}
\centerline{
\includegraphics[width=4.5cm,angle=90]{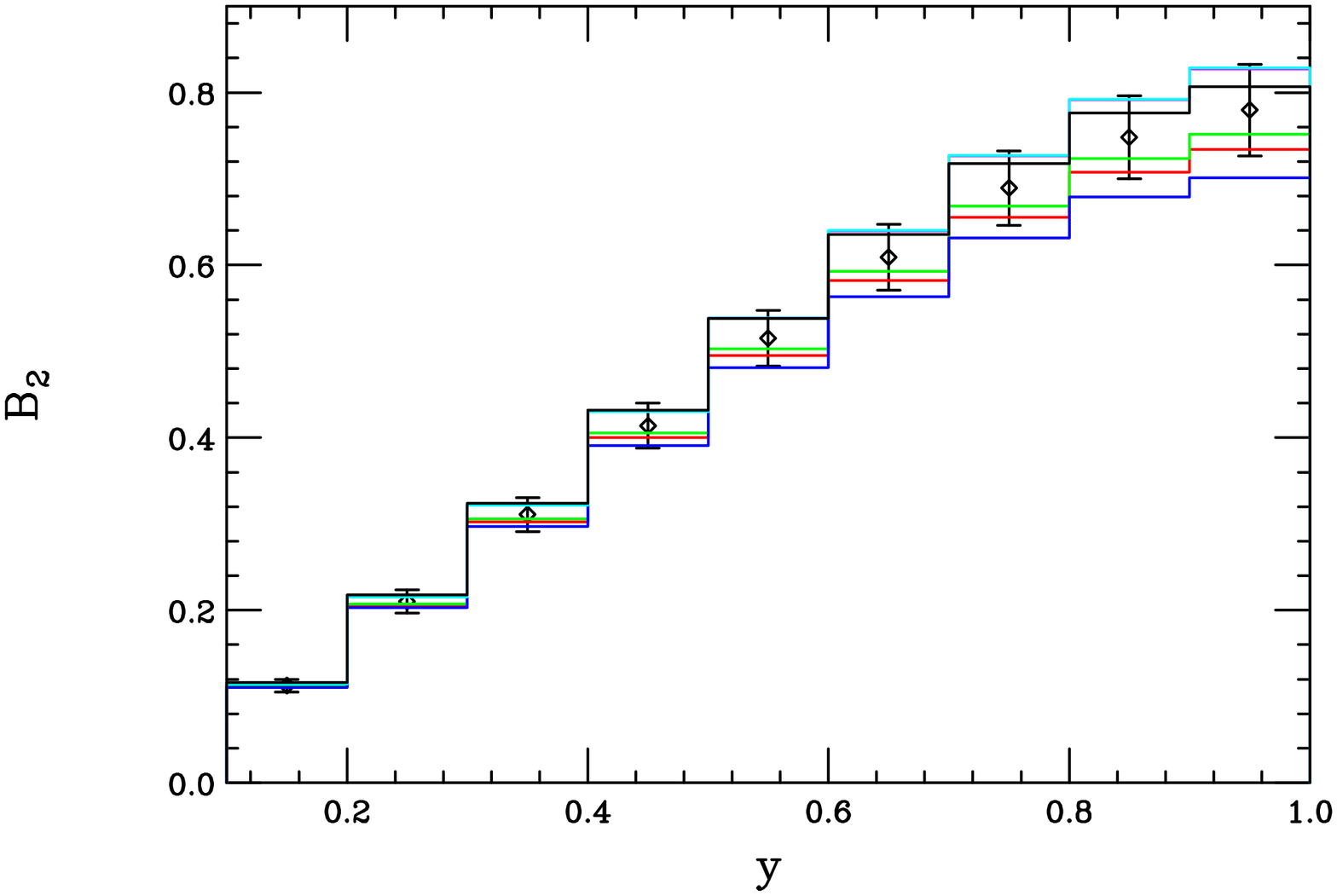}}
\caption{Same as the previous figure but now for the asymmetries $C_R,B_{1,2}$.  The histograms, from top to bottom on the right-hand side 
of the top-left (top-right), [bottom] panel correspond to $Z'$ models LRM, SSM, $\chi$, $\eta$, $\psi$, ALRM
(SSM, LRM, $\chi$, $\eta$, $\psi$, ALRM),
[ALRM, LRM, SSM, $\chi$, $\psi$, $\eta$], respectively.}
\label{fig4}
\end{figure}

An example of these $ep$ asymmetries probing the $Z'$ couplings can be found in Figs.~\ref{fig3} and ~\ref{fig4}. Here we see that the various 
asymmetries show quite different sensitivities to different possible $Z'$ models. To obtain these results we required that $0.25 \leq x$ and 
$0.1 \leq y$ to remove the SM-dominated low-$Q^2$ region and we have assumed $80\%$ beam polarization. As $\sqrt s$ increases with fixed 
luminosity there is some increase in the sensitivity to heavier $Z'$ in these asymmetries but this is partly offset by the diminishing cross 
sections that result in lower statistics. This implies that measurements at the LHeC will only be useful for $Z'$ coupling determinations 
provided that the $Z'$ is not much heavier than $\simeq 1.5$ TeV if it has canonical strength couplings as in the sample cases above. A complete 
evaluation of the LHeC's capabilities, however, must await a more detailed study of the proposed properties of this collider{\cite {lhec}}.


\begin{acknowledgments}

Work supported by Department of Energy contract DE-AC02-76SF00515.
\end{acknowledgments}

\end{document}